# Design Simulation of Czerny-Turner Configuration-based Raman Spectrometer using physical optics propagation algorithm

*Muddasir Naeem[1], Noor-ul-ain Fatima[1], Mukhtar Hussain[2, a], Tayyab Imran[3,b], Arshad Saleem Bhatti[1]*

[1]Department of Physics, COMSATS University Islamabad, Park Road, 45550, Islamabad, Pakistan.
[2]GoLP/Instituto de Plasmas e Fusão Nuclear-Laboratório Associado, Instituto Superior Técnico, Universidade de Lisboa, 1049-001 Lisboa, Portugal.
[3]Research Laboratory of Lasers (RLL)-Group of Laser Development (GoLD), SBASSE, Department of Physics, Lahore University of Management Sciences, Lahore, Punjab 54792, Pakistan.

**Abstract**
We report the design simulation of the Raman spectrometer using optical system design software Zemax. The design is based on the Czerny-turner configuration which includes an optical system consisting of an entrance slit, two concave mirrors, reflecting type diffraction grating, and an image detector. The system's modelling approach is suggested by introducing the corresponding relationship between detector pixels and wavelength, linear CCD receiving surface length, and image surface dimension. Simulations have been carried out using the POP (Physical Optics Propagation) algorithm. Spot diagram, relative illumination, irradiance plot, Modulation Transfer Function (MTF), geometric, and encircled energy simulated for designing the Raman spectrometer. The simulation results of the Raman spectrometer's using a 527 nm wavelength laser as an excitation light source are presented. The present optical system is designed in sequential mode and a Raman spectrum observed in the range of 530 to 630 nm. The analysis shows that the system's image efficiency is higher, predicting that it is possible to build an efficient and cost-effective Raman spectrometer for optical diagnostics.
**Keywords:** Raman spectrometer, Zemax simulations, POP Algorithm

1. **INTRODUCTION**

Spectrometry is a general term used for analyzing the specific spectrum and employed for the analysis of materials. Raman spectrometry [1] is used to study molecular structures and identify molecules in unknown samples and give spectral information [2]. Jin Xinghuan designed the modern grating-based Raman spectrometer. The spectral resolution, wavelength range, and the simple spectrometer's structural parameters are the contemporary way of thinking geometrical models [3].

Different designs of the Raman spectrometer have been published. Wang C, Chen H designed a crossed Czerny–Turner spectrometer using convergent illumination of the grating for the fluorescence spectrum of the organic particles [4]. Yinchao Zhang, Chen Wang proposed a double-grating monochromator with a different fiber arrangement [5]. The Offner system with a convex grating is exceptional in aberration correction and distortion control [5,6]. Czerny and Turner first showed that the coma aberration introduced by the off-axis reflection from a spherical mirror could be corrected by symmetrical, but oppositely oriented, a spherical mirror in spectrometer design [7]. The Czerny-Turner imaging spectrometer, a plane grating, and two spherical mirrors are configured in a coma-free geometry with the Shafer equation satisfied used to resolve spectral intensity [8].

This article introduced the design simulation of the Raman spectrometer's optical structure using plain, reflecting grating and a focusing spherical mirror, shortening the overall system's length to obtain higher resolution. Thus enabling the development of robust and low-cost Raman spectrometer for optical measurements.

## 2. DESIGN PARAMETERS

The lens data editor window in Zemax [9], shown in Table 1, defines the different designing parameters, such as the surface type of each lens, the radius of curvature, thickness, and focal length. We adopted the physical optics propagation (POP) algorithm over geometrical ray-tracing because geometrical ray-tracing can only be used when diffraction limits are negligible.

|  | Surface Type | Comment | Radius of Curvature (mm) | Thickness (mm) | Glass | Semi-Diameter (mm) |
|---|---|---|---|---|---|---|
| **OBJ** | Standard | Source | Infinity | 0.90 |  | 0.00 |
| 1 | Standard | Sample | Infinity | 0.90 |  | 0.30 |
| 2 | Standard | Slit | Infinity | 4.50 |  | 0.08 |
| 3 | Standard | Collimating mirror | -10.00 | -3.00 | MIRROR | 0.61 |
| 4 | Diffraction Grating | Grating | Infinity | 3.00 | MIRROR | 0.52 |
| 5 | Standard | Focusing mirror | -10.00 | -4.50 | MIRROR | 0.61 |
| 6 | Standard | Detector | Infinity | - |  | 0.30 |

**Table 1.** Lens data editor window in Zemax

The system includes an entrance slit, spherical collimating, and focusing mirrors, grating, and CCD (Charge-coupled device) detector [10]. Diffraction gratings are exceptional due to their imaging efficiency. However, they are usually unable to achieve a flat focus curve [11], so a spherical mirror is needed to concentrate the subject on the image surface. Figure 1A shows the schematic diagram, which is the layout for the Raman spectrometer's optical configuration. Raman scattered light from the sample is focused on the entrance slit with the focusing lens of numerical aperture 0.16 which is further collimated with collimating mirrors and incident on the diffraction grating. The Raman signal diffracted from the grating and split up into its components which is focused on the CCD camera through the focusing mirror (Figure. 1A). The radius of the curvature of the collimating and focusing mirror is $R_1$ and $R_2$, respectively.

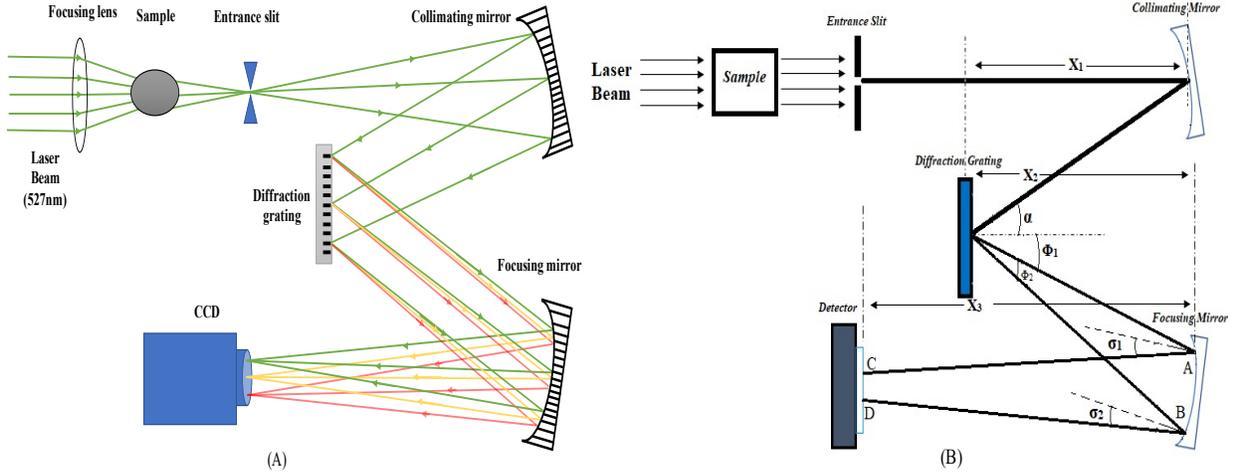

**Figure 1. A.** Layout of Raman Spectrometer. **B.** Analysis of the imaging plane in Raman spectrometer

A sample is defined before the entrance slit of the spectrometer. The upper and the lower energy levels of the sample have certain vibrational levels. We choose a 527 nm centre wavelength laser as the light source, whose energy is equal to the energy bandgap between the upper and lower energy level of the sample. The sample absorbs the incident light and excites to the upper energy level. Electrons in the excited state decay to a lower vibrational level through a non-radiative process, and then decay to a lower energy level and emit photons of different energy, which is called the Raman spectrum. The Raman spectrum of the sample is distributed usually in the range of 200 ~ 2500 cm$^{-1}$ [3], and the Raman spectra wavelength range observed in the experiment is from 530 nm to 630 nm. A 1024 ×1024 pixels charge-coupled device (CCD) is chosen to record the spatial profile. The scale of each pixel is ( 5 × 5 )μm, and the receiving surface of CCD is 1.25 cm$^2$. The numerical aperture (NA) of the object side is chosen to be 0.16, while the grating constant is 0.100 lines/μm, and the width of the slit is 30 μm. We have chosen the size of the CCD which is comparable to the image plane [12].

The initial parameters of the system design are shown in Table 2.

| Parameter | Value |
|---|---|
| Laser Wavelength | 527 nm |
| Observed Raman Spectra | (530 – 630) nm |
| Entrance Pupil Diameter | 0.16 |
| Grating Constant | 0.100 lines/μm |
| CCD Pixel | 1024 x 1024 |
| Slit width | 30 μm |
| Size of Pixel | (5 x 5) μm |
| $R_1$ | -10 mm |
| $R_2$ | -10 mm |
| $x_1$ | 3 mm |
| $x_2$ | 3 mm |
| $x_3$ | 4.5 mm |

**Table 2.** Initial structure parameters of Raman spectrometer

In Figure 1B, angles $\phi_1$ and $\phi_2$ are the diffraction angles of the two edge wavelengths after passing through the diffraction grating; they intersect the focusing mirror at A and B. $\sigma_1$, and $\sigma_2$ are the angles between the reflected light of the two edge wavelengths on the focusing mirror and the horizontal line. Finally, they gather on points C and D of the linear CCD.

Using the diffraction grating formula,

$$d * sin\theta = n\lambda \quad (1)$$

Where $d$ is the grating constant, diffraction angles $\phi_1$ and $\phi_2$ can be calculated as:

$$d * (\sin(\alpha) + \sin(\phi_1)) = n\lambda_1 \quad (2)$$
$$d * (\sin(\alpha) + \sin(\phi_2)) = n\lambda_2 \quad (3)$$

Taking the diffraction order equal to 1 (n=1). Minimum size (length) of the focusing mirror to reflect the two edge wavelengths according to the geometrical structure:

$$AB = x_2 * [\tan(\alpha + \phi_1) - \tan(\alpha + \phi_2)] \quad (4)$$

After reflection from the focusing mirror, the horizontal angle of the diffracted rays is changed by $2\theta$ as given below:

$$\sigma = \alpha + \phi - 2\theta \quad (5)$$

The size of the image is calculated using the geometrical structure:

$$CD = AB - x_3 * [\tan(\sigma_2) - \tan(\sigma_1)] \quad (6)$$

Equation 6 shows the minimum length of the CCD detector surface. Based on the original configuration and using the equations (1- 6), all the necessary parameters to design the Raman spectrometer can be determined.

## 3. SIMULATION RESULTS AND ANALYSIS

After optimization, $x_1$ increases to 3 mm. The light diffracted by the grating has a certain convergence angle, corresponding to $x_3$ is reduced to 4.5 mm, other parameters like detector size and pixels, and mirror focal length unchanged. The optimized spot diagram of the adjacent wavelengths is shown in Figure 2A. An image formed on the spectrometer's image plane shows that the adjacent wavelengths at different positions are well separated and distinguished.

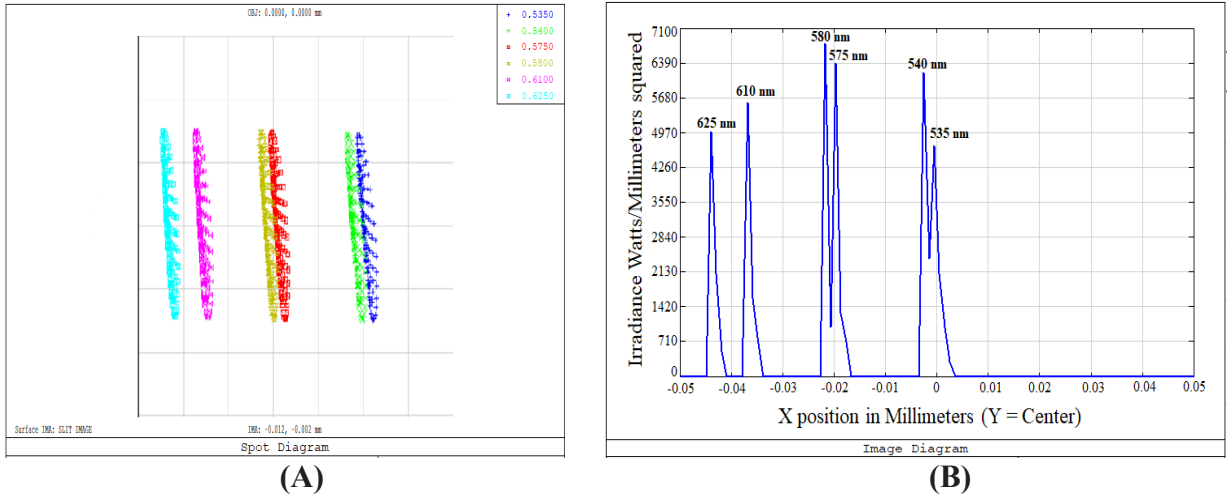

(A)          (B)

**Figure 2.** **A.** Spot diagram of 535 nm (blue), 540 nm (green), 575 nm (red), 580 nm (gold), 610 nm (purple) & 625 nm (sky blue). **B**. Intensity distribution at various wavelengths 535 nm, 540 nm, 575 nm, 580 nm, 610 nm, 625 nm

It can be seen from Figure 2B that the energy spectrum of the neighbouring wavelengths at various wavelengths can be easily distinguished. The MTF diagram at the centre wavelength (575 nm) is shown in Figure 3, which gives the contrast ratio between the input and output images. The MTF curve is approximately the same in the whole band, so the only curve of the centre wavelength is given here. MTF shows how the spatial frequency content of the entity is correctly transferred to the image and describe the performance of the optical system. The higher the value of MTF, the greater will be the image quality of the device. As shown in Figure 3, when the spatial frequency is 10 per mm$^{-1}$, the optical transfer function (OTF) is greater than 0.8, which means that the efficiency of the designed spectrometer system is higher [13].

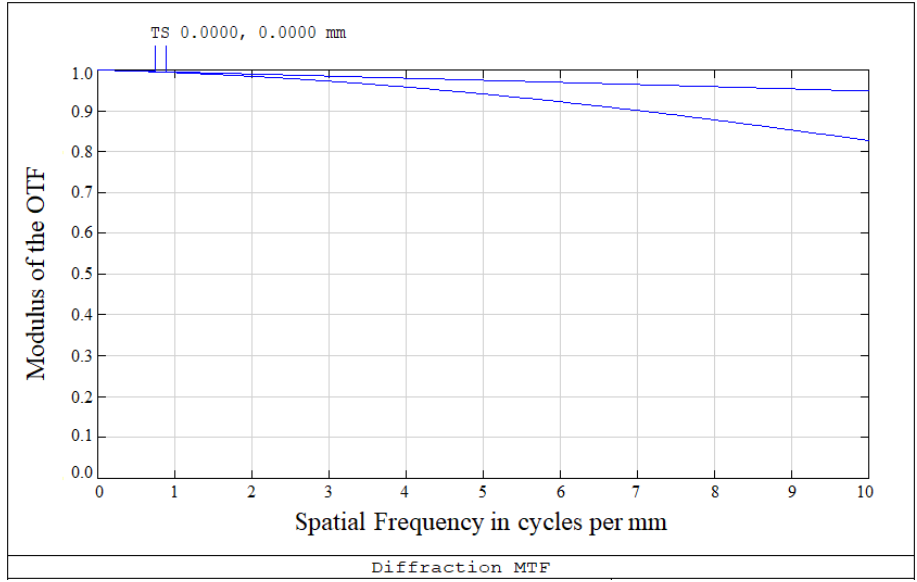

**Figure 3.** MTF curve at 575 nm

All results are obtained in the Sequential Mode of Zemax. In this mode, the slit width can be measured only by increasing the height of the point light source. Follow the above settings to trace the light, and the results obtained in this mode are closer to the real-world observations.

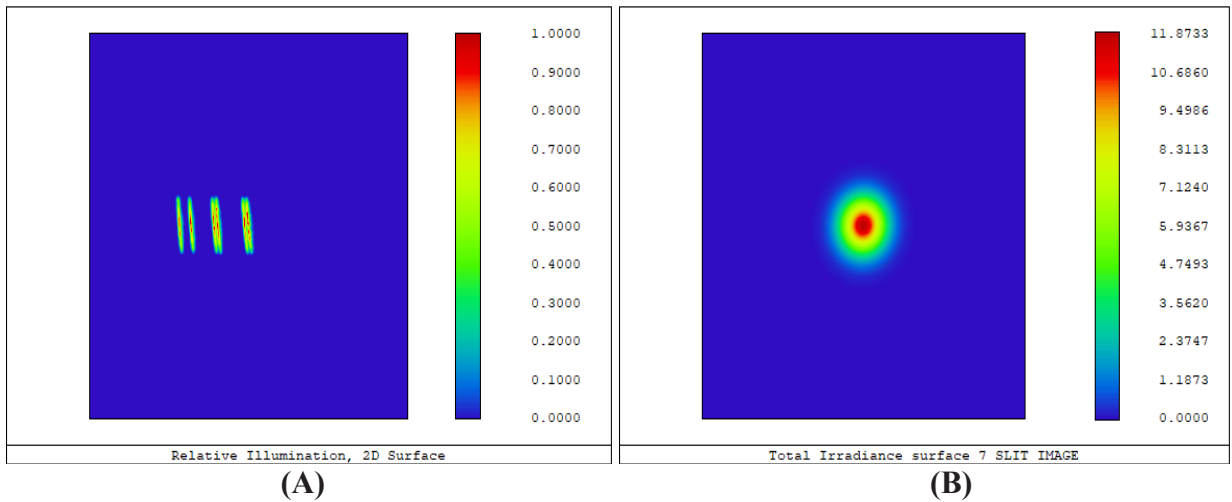

**Figure 4.** **A.** Wavelength band imaged on the plane of the detector. **B.** Irradiance and beam spot at the detector surface

The intensity distribution of adjacent wavelengths shown in Figure 4A shows the intensity distribution of all the different wavelengths expressed by the geometric linear response function. The linear response function is the cross-sectional representation of the image density distribution [14]. The length and width of the receiving surface of the detector are 10 mm and 5 mm, respectively. The wavelength band imaged on the detector, and the images produced by the two edge rays are similar to the two ends of the detector (Figure 4A ). Using the CCD, the light of the neighbouring wavelengths of separate bands can also be separated, which indicates that the illumination of the adjacent wavelengths is easily separable and distinguishable with the present detector pixels. This finding further shows the functional viability of this system. The radiant flux received by the detector surface per unit area and beam spot size at the detector surface for 580 nm wavelength is shown in Figure 4B

$$I \; \alpha \; 1/\lambda^4$$

Illuminations at longer wavelengths result in a decrease in Raman signals. There is less illumination for a higher wavelength (total irradiance) than the lower wavelength in the Raman spectrometer, so the above relation is working correctly in our designed Raman spectrometer. The optical term encircled energy which describes the energy concentration in the optical picture or predicted laser in a defined area. We have observed the energy concentration variations concerning change in the size of the image on the detector; as the point image's size increases, the fraction of enclosed energy increases which is consistent as reported [15].

## 4. CONCLUSIONS

The Raman spectrometer utilizing diffraction grating incorporates the standard Czerny-Turner system features with the flat-field grating spectrometer system. The spectrometer system's modelling approach is suggested by introducing the corresponding relationship between pixel and wavelength, linear CCD receiving surface length, and image surface dimension. The excitation light with a centre wavelength of 527 nm is used for simulation and optimization in the Zemax optical modelling program. Spot diagram, irradiance plot, MTF, and geometric encircled energy for Raman Spectra of the sample, which lie in the visible range, are simulated. The analysis shows that the image efficiency of the system is higher, and Raman spectra can be obtained from a sample using the above designed Raman spectrometer system. This study could pave the way for the development of a robust, miniature and compact low-cost Raman spectrometer for optical measurements.